\documentclass[3p,times]{elsarticle}

\usepackage{ecrc}

\usepackage{epstopdf}

\newcommand{\beq}{\begin{equation}}
\newcommand{\eeq}{\end{equation}}
\newcommand{\beqa}{\begin{eqnarray}}
\newcommand{\eeqa}{\end{eqnarray}}
\newcommand{\bseq}{\begin{subequations}}
\newcommand{\eseq}{\end{subequations}}

\def\x{{\boldsymbol x}}

\def\k{{\boldsymbol k}}
\def\q{{\boldsymbol q}}
\def\p{{\boldsymbol p}}

\def\0{{\boldsymbol 0}}

\def\F{{\boldsymbol F}}

\def\kk{{\kappa}}
\def\pp{{\hat{p}}}

\def\lsim{\raise0.3ex\hbox{$<$\kern-0.75em\raise-1.1ex\hbox{$\sim$}}}
\def\gsim{\raise0.3ex\hbox{$>$\kern-0.75em\raise-1.1ex\hbox{$\sim$}}}

\volume{00}

\firstpage{1}

\journalname{Nuclear Physics A}

\runauth{Andrea Beraudo}

\jid{nupha}

\jnltitlelogo{Nuclear Physics A}

\usepackage{graphicx}
\usepackage{amsmath,amssymb}

\begin{document}

\begin{frontmatter}



\title{Dynamics of heavy flavor quarks in high energy nuclear collisions}

\author{Andrea Beraudo}
\address{Istituto Nazionale di Fisica Nucleare - Sezione di Torino\\
Via Pietro Giuria 1, 10125 Torino}




\begin{abstract}
A general overview on the role of heavy quarks as probes of the medium formed in high energy nuclear collisions is presented. Experimental data compared to model calculations at low and moderate $p_T$ are exploited to extract information on the transport coefficients of the medium, on possible modifications of heavy flavor hadronization in a hot environment and to provide quantitative answers to the issue of kinetic (and chemical, at conceivable future experimental facilities) thermalization of charm. Finally, the role of heavy flavor at high $p_T$ as a tool to study the mass and color-charge dependence the jet quenching is also analyzed.    
\end{abstract}


\end{frontmatter}



\section{Introduction}\label{intro}
Heavy flavor quarks play a peculiar role as probes of the medium formed in high energy nuclear collisions. If soft observables are nicely reproduced by hydrodynamics, assuming to deal with a system at local thermal equilibrium (no matter why), and jet quenching is interpreted in terms of the energy degradation of high-$p_T$ partons playing the role of external probes, the description of heavy flavor observables requires to develop a setup allowing one to deal with the more general situation of particles which would asymptotically approach kinetic equilibrium with the background medium: such a tool is represented by transport calculations, which I'm going to discuss in my contribution. In the last part of the paper I will also address the quenching of heavy flavor jets, allowing one to study the mass and color charge dependence of the parton energy loss.

Before moving to phenomenology one should first of all answer the question why charm and beauty are considered heavy: the reasons are at least three. First of all their mass $M$ is much larger then $\Lambda_{\rm QCD}$, so that their initial production is a hard process described by pQCD. Secondly $M\!\gg\! T$, making thermal production during the limited lifetime of the plasma negligible: charm and beauty multiplicity is set by the initial hard processes. Finally $M\!\gg\! gT$, $gT$ being the typical momentum exchange with the plasma particles, entailing that many soft scatterings are necessary to change significantly the momentum/trajectory of the heavy quarks; notice that for realistic temperatures $g\!\sim\! 2$, so that one might wonder whether charm has to be considered really ``heavy'', at least at the beginning of the fireball evolution.

\begin{figure}
\begin{center}
\includegraphics*[height=5.2cm]{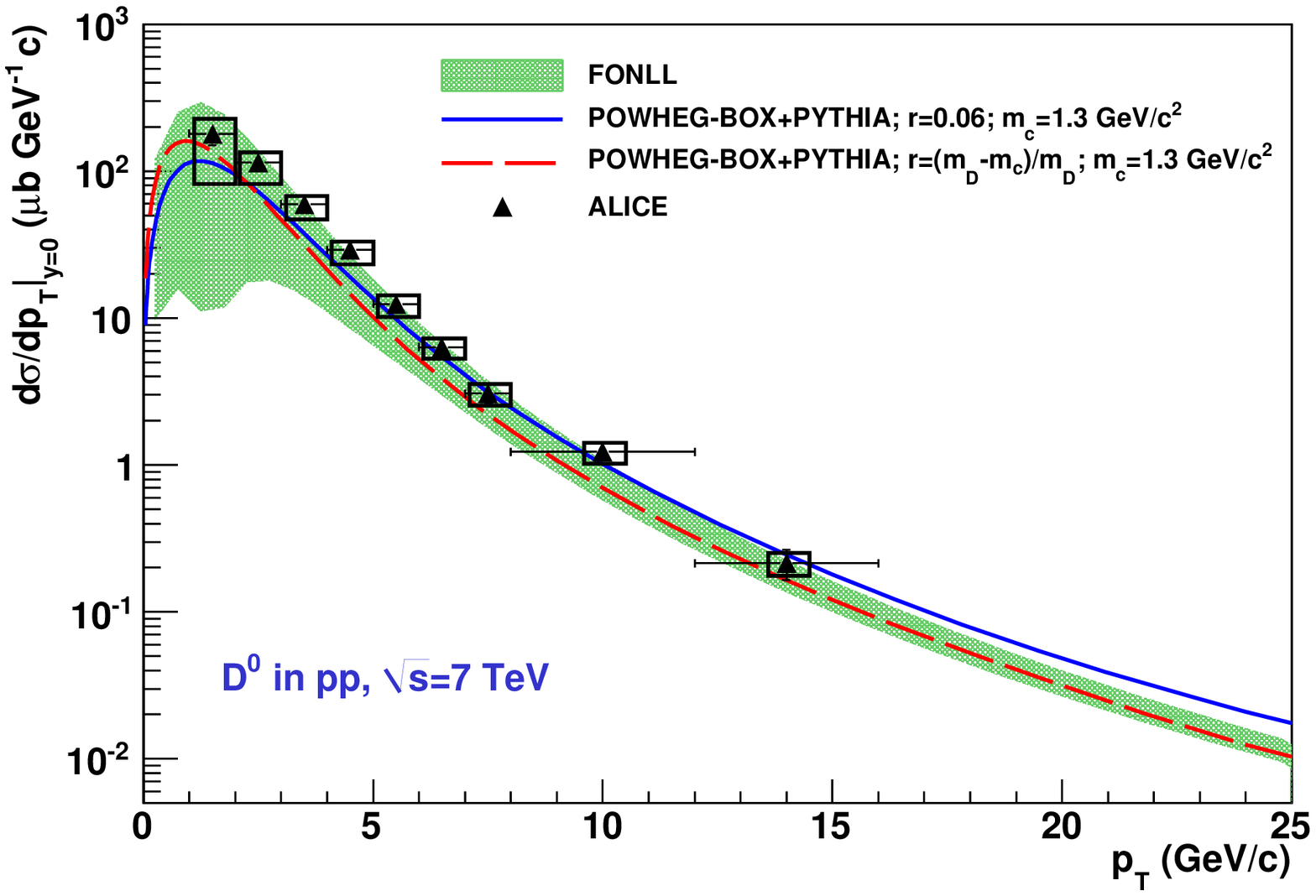}
\includegraphics*[height=5.2cm]{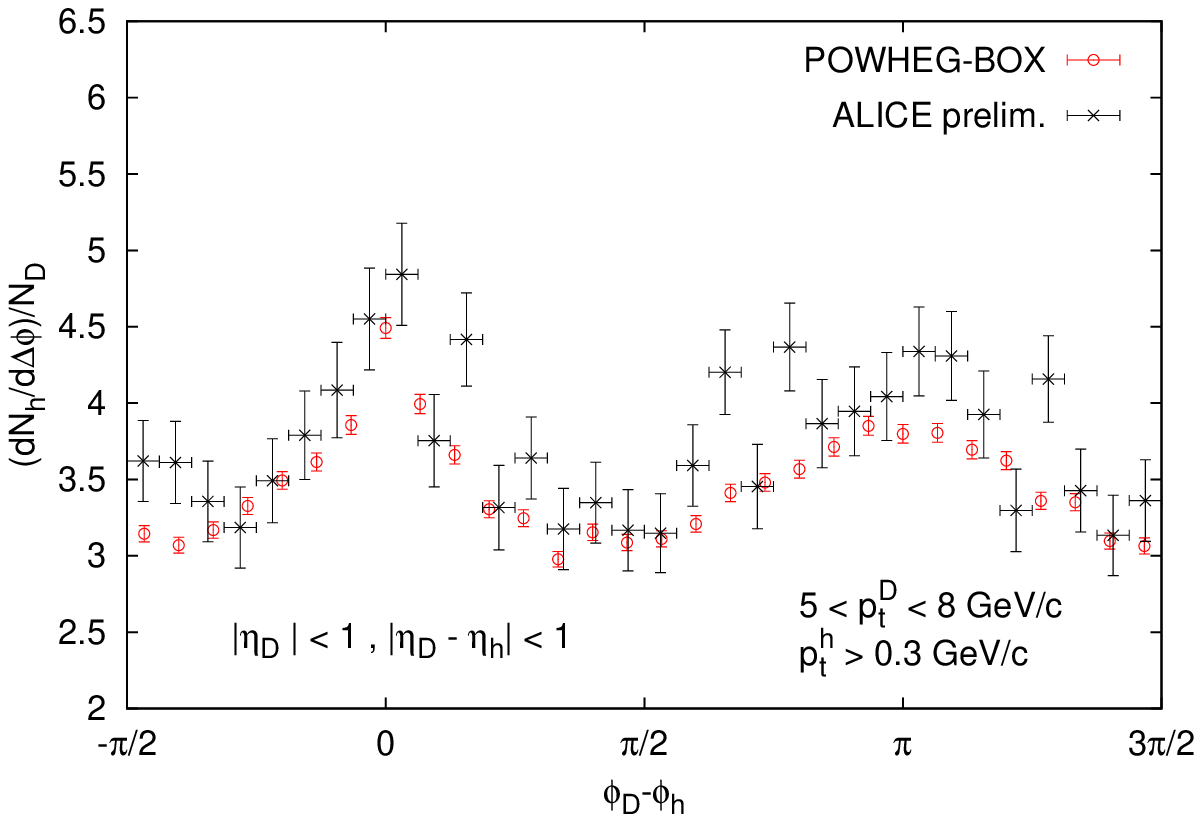}
\caption{Left panel: differential cross section for $D^0$ production in p-p collisions at $\sqrt{s}=7$ TeV; predictions of the FONLL calculation and of the POWHEG-BOX event generator (with different charm fragmentation functions) are compared to ALICE data. Right panel: POWHEG-BOX predictions for D-hadron correlations in p-p collisions at $\sqrt{s}=7$ TeV~\cite{inprogress} compared to preliminary ALICE results~\cite{sandro}.}
\label{fig:ppbenchmark}
\end{center}
\end{figure}
The state of the art in the description of the initial $Q\overline{Q}$ production is represented by NLO pQCD calculations (POWHEG, MC@NLO) for the hard process interfaced to some event generators (PYTHIA, HERWIG) to simulate the Initial and Final State Radiation and other non perturbative processes (intrinsic $k_T$, Underlying Event and hadronization). One can employ an automated tool like the POWHEG-BOX package to perform such calculations getting a fully exclusive information of the final state. As it can be seen in Fig.~\ref{fig:ppbenchmark} this allows one to satisfactory describe heavy flavor production in p-p collisions, reproducing not only their inclusive spectra, but also (letting PYTHIA take care of the hadronization of the whole event) more differential observables like D-h azimuthal correlations, currently at the center of important experimental efforts. A systematic comparison of the outcomes of the various pQCD event generator and of other automated tool like the FONLL calculation can be found in~\cite{pQCDtools}.

\section{Transport, flow and thermalization in the QGP}
In this section I will present a critical overview of the transport calculations used to describe the dynamics of heavy quarks in the medium formed in heavy-ion collisions: the comparison with the experimental data should ideally allow one to put tight constraints on the transport coefficients of the QGP. A more ambitious approach consists in deriving the latter directly from the QCD Lagrangian, either by weak-coupling calculations or through non-perturbative approaches like lattice-QCD (lQCD) simulations, whose recent results will be discussed. There are important issues to address concerning the fate of heavy flavor in heavy-ion collisions: how close are heavy quarks to thermalization? Are the final hadronic observables able to answer this question? what could be the role of experiments at increasing $\sqrt{s_{\rm NN}}$? In the following the discussion will touch all the above issues.

\begin{figure}
\begin{center}
\includegraphics*[width=0.39\textwidth]{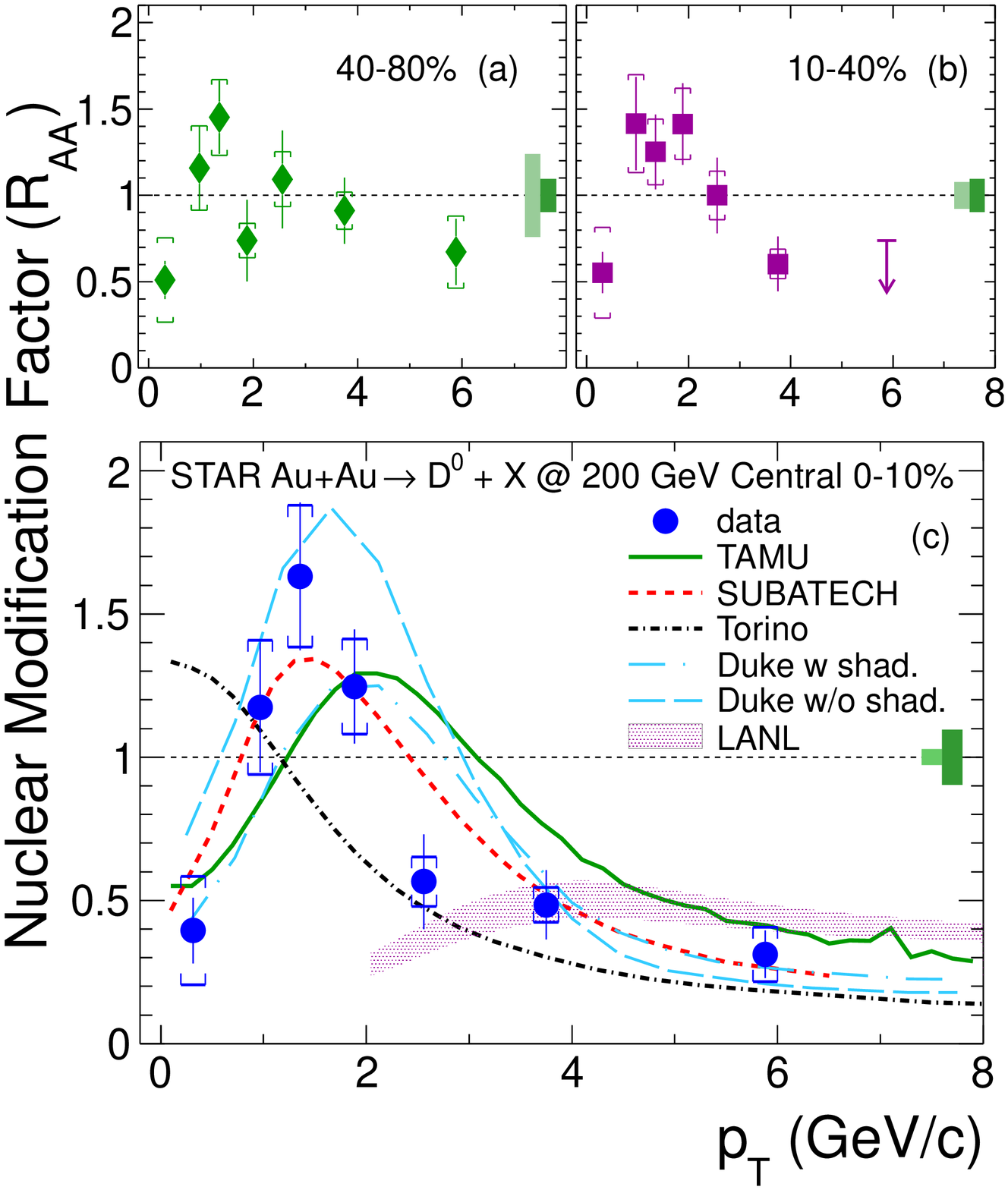}
\includegraphics*[width=0.48\textwidth]{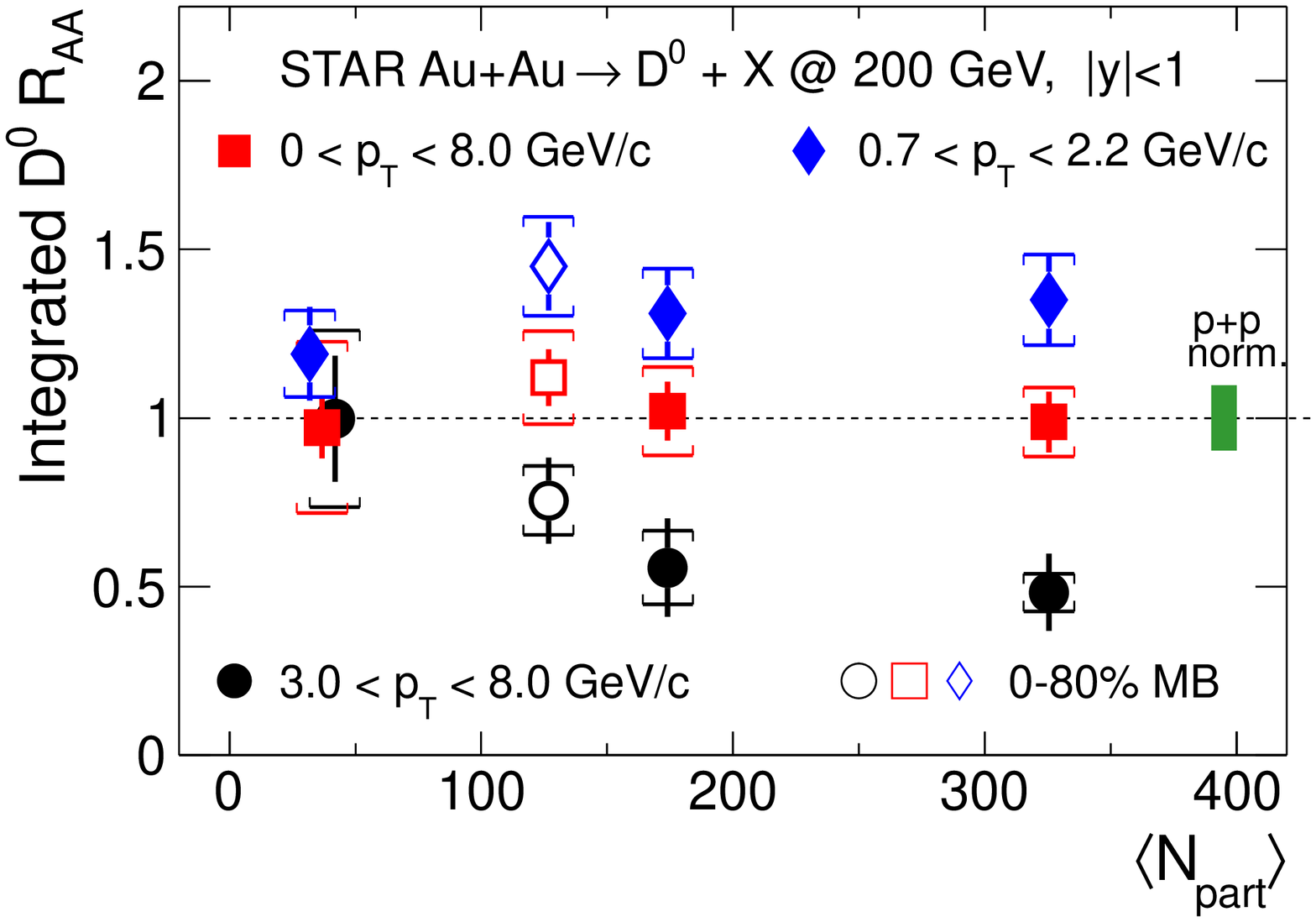}
\caption{Left panel: the nuclear modification factor of $D^0$ mesons in Au-Au collisions at RHIC for various centrality classes. STAR data~\cite{STAR} in the $0-10\%$ more central events are compared to theory predictions. Right panel: integrated $R_{AA}$ over various $p_T$ windows as a function of centrality. The flat behavior around 1 of the data integrated down to zero momentum confirms the hypotesis of an initial hard $Q\overline{Q}$ production scaling with the number of binary nucleon-nucleon collisions.}
\label{fig:STAR-RAA}
\end{center}
\end{figure}

The starting point of all transport calculation is the Boltzmann equation for the evolution of the heavy quark phase-space distribution 
\beq
{\frac{d}{dt}f_Q(t,\x,\p)=C[f_Q]}\quad{\rm with}\quad C[f_Q]=\int d\k[{w(\p+\k,\k)f_Q(t,\x,\p+\k)}-{w(\p,\k)f_Q(t,\x,\p)}]
,\label{eq:Boltzmann}
\eeq
where the collision integral $C[f_Q]$ is expressed in terms of the $\p\to\p-\k$ transition rate $w(\p,\k)$
The direct solution of the Boltzmann equation in the evolving inhomogeneous fireball produced in heavy-ion collisions is challenging; results can be found in~\cite{Pol,BAMPS}. However as long as $k\ll p$ ($k$ being typically of order $gT$) one can expand the collision integral in powers of the momentum exchange. Truncating the expansion to second order one gets the Fokker-Planck (FP) equation (in the following the dependence on $\x$ and on possible mean-fields $\F$ will be neglected)
\beq
\frac{\partial}{\partial t}f_Q(t,\p)=\frac{\partial}{\partial p^i}\left\{{A^i(\p)}f_Q(t,\p)+\frac{\partial}{\partial p^j}[{B^{ij}(\p)}f_Q(t,\p)]\right\},
\eeq
expressed in terms of the friction and momentum broadening coefficients $A(p)$ and $B_{0/1}(p)$ defined as
\beq
{A^i(\p)}\equiv\int d\k\, k^iw(\p,\k)\equiv{A(p)}\,p^i\quad{\rm and}\quad{B^{ij}(\p)}\equiv\frac{1}{2}\int d\k\, k^ik^jw(\p,\k)\equiv\hat{p}^i\hat{p}^j{B_0(p)}+(\delta^{ij}-\hat{p}^i\hat{p}^j){B_1(p)}.
\eeq
The integro-differential equation~(\ref{eq:Boltzmann}) has been reduced then to a standard partial differential equation, easier to solve. The need to follow the propagation of each heavy quark inside an inhomogeneous medium subject to a hydrodynamic evolution suggests to recast the FP equation into an equivalent form more suited to numerical simulations: the relativistic Langevin equation. The latter belongs to the class of stochastic differential equations and represents the approach adopted in most of the theoretical studies. In its discretized form
\beq
{\Delta \vec{p}}/{\Delta t}=-{\eta_D(p)\vec{p}}+{\vec\xi(t)},\label{eq:Langevin}
\eeq
it provides a recipe to update the heavy quark momentum through the sum of a deterministic friction force and a random noise term specified by its temporal correlator
\beq
\langle\xi^i(\p_t)\xi^j(\p_{t'})\rangle\!=\!{b^{ij}(\p_t)}{\delta_{tt'}}/{\Delta t}\qquad{b^{ij}(\p)}\!\equiv\!{\kk_\|(p)}\pp^i\pp^j+{\kk_\perp(p)}(\delta^{ij}\!-\!\pp^i\pp^j)
\eeq
After evaluating the transport coefficients $\kappa_{\|/\perp}(p)$ (representing the average longitudinal/transverse squared momentum exchanged with the plasma per unit time) and $\eta_D(p)$ (the latter being fixed by the Einstein relation, so that particles asymptotically reach kinetic equilibrium) one has then to solve Eq.~(\ref{eq:Langevin}) throughout the whole medium evolution. At a given critical temperature heavy quarks are then hadronized and the final particle spectra (of $D$ mesons, heavy-flavor electrons, $J/\psi$'s from $B$ decays...) can be compared to the experimental data. As an example of recent results, in Fig.~\ref{fig:STAR-RAA} we show STAR data~\cite{STAR} for the $R_{AA}$ of $D^0$ mesons compared to the predictions of various transport calculation~\cite{rapp,gossiaux,torino,bass}. The main feature of STAR data is represented by the sharp peak around $p_T\sim 1.5$ GeV, whose origin will be discussed in more detail in the following: for the moment one can notice that the models displaying a better agreement with the data are the ones modeling the hadronization of the heavy quarks through the coalescence with light thermal partons, whose radial flow tends to boost very soft charm quarks towards larger $p_T$.

\begin{figure}
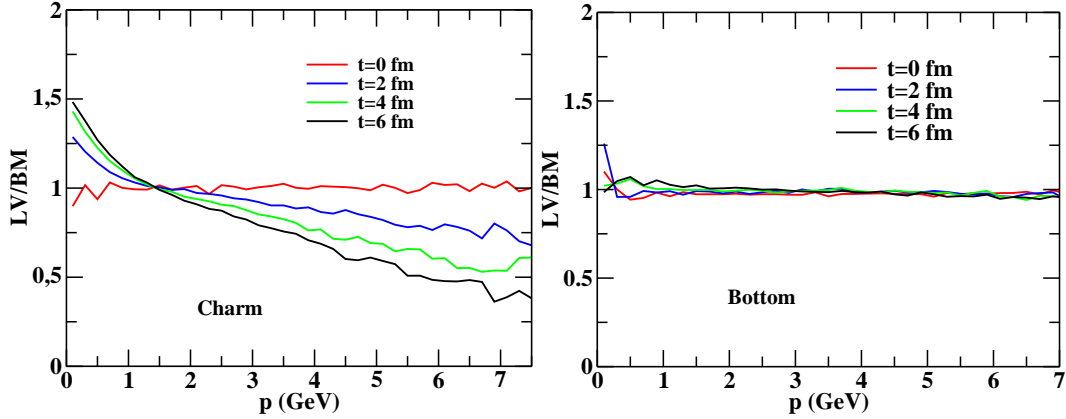

\begin{center}
\includegraphics*[width=0.42\textwidth]{LV_BM_c.eps}
\includegraphics*[width=0.42\textwidth]{LV_BM_b.eps}
\caption{Ratio of the charm (left panel) and beauty (right panel) spectra obtained with a Langevin equation with respect to the result of the full Boltzmann equation for various values of time. The medium is assumed to be a plasma at $T\!=\!0.4$ GeV and a Debye mass $m_D\!=\!0.83$ GeV is employed in the calculations. With this choice of parameters deviations between the two approaches are present in the case of charm, while for beauty the Langevin equation is always an excellent approximation.}
\label{fig:BvsLang}
\end{center}
\end{figure}

As we have seen most of the heavy flavor transport calculations are based on the Langevin equation which, although being a very convenient numerical tool allowing one to establish a link between observables and QCD transport coefficients, is nevertheless based on a soft-scattering expansion ($k\ll p$) of the collision integral $C[f_Q]$ truncated at second order, summarizing the effects of the medium into a friction and a momentum diffusion term. This condition may be not always fulfilled, in particular for charm, hence the interest in comparing the results obtained solving the full Boltzmann equation with the ones of its corresponding Langevin limit. Such an analysis was performed in~\cite{greco} and results are shown in Fig.~\ref{fig:BvsLang}. As it can be seen, for realistic values of the temperature and of the Debye mass, the Langevin results display deviations from the exact solution of the Boltzmann equation for charm, while in the case of beauty the agreement between the two approaches is always perfect.

\begin{figure}
\begin{center}
\includegraphics*[height=4.9cm]{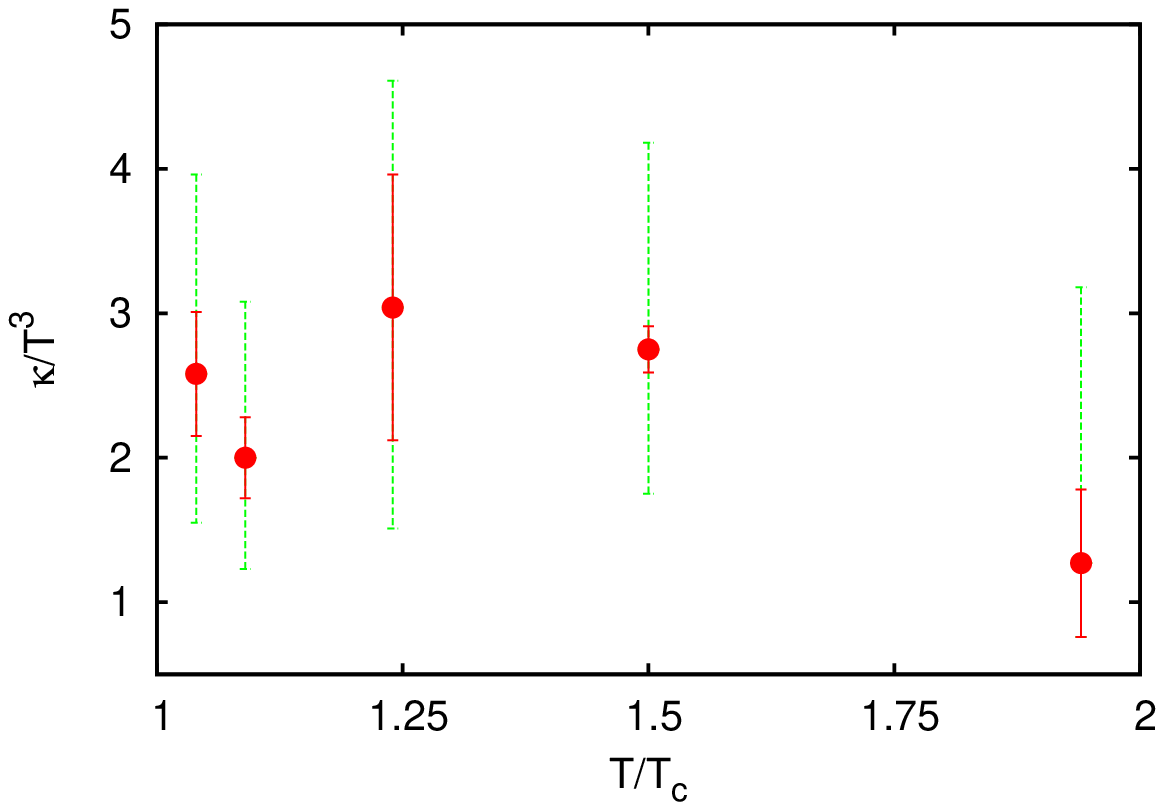}
\includegraphics*[height=4.9cm]{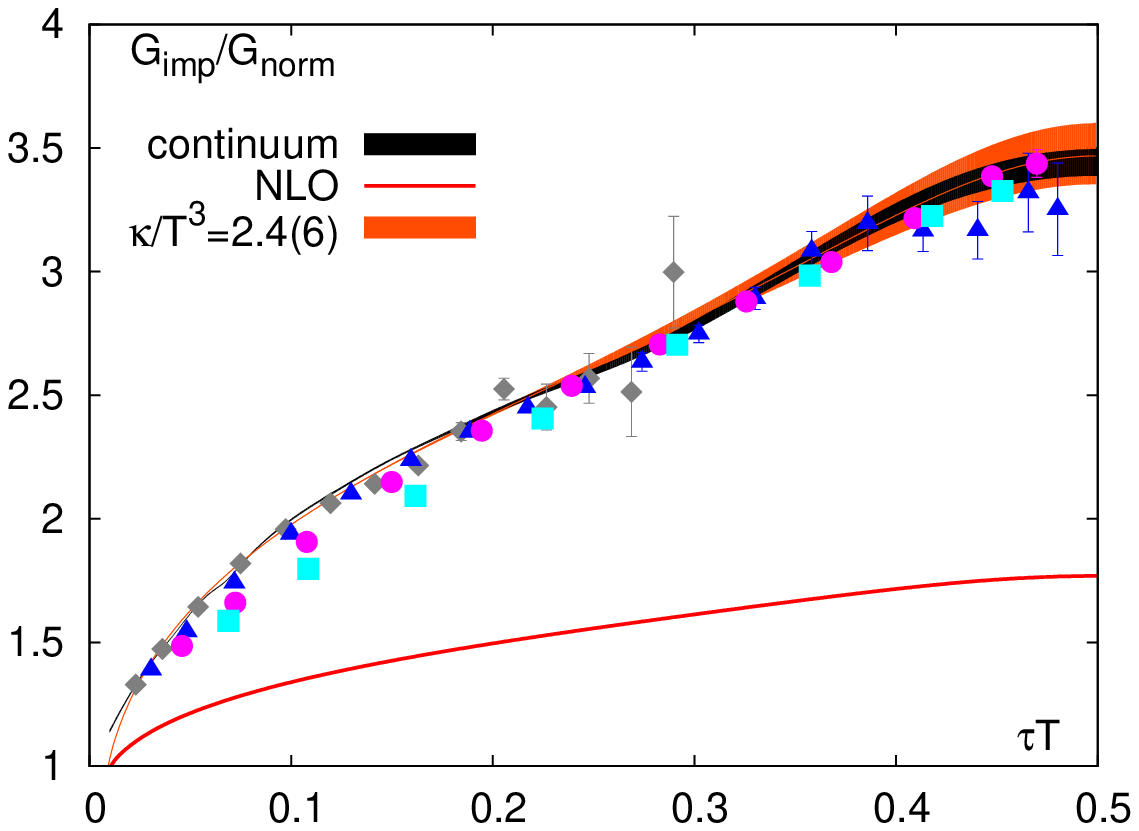}
\caption{Lattice-QCD results~\cite{lat1} for the heavy quark momentum diffusion coefficient $\kappa$ in a gluon plasma as a function of the temperature (left panel) and a study of its continuum limit (right panel) from Ref.~\cite{lat3}.}
\label{fig:lQCD}
\end{center}
\end{figure}

Among the approaches to evaluate heavy quark transport coefficients, lattice-QCD calculations represent an important tool, having the advantage of providing a non-perturbative answer, not limited to a weak-coupling regime. Current results refer to the limit of an infinitely heavy quark, assumed to be described by the Langevin equation~(\ref{eq:Langevin}), in which one can neglect the momentum dependence of the strength of the noise (the quark being at rest), so that
\beq
\langle\xi^i(t)\xi^j(t')\rangle\!=\!\delta^{ij}\delta(t-t'){\kappa}\quad\longrightarrow\quad{\kappa}=\frac{1}{3}\int_{-\infty}^{+\infty}\!\!dt\langle\xi^i(t)\xi^i(0)\rangle_{\rm HQ}
\underset{p\to 0}{\sim}\frac{1}{3}\int_{-\infty}^{+\infty}\!\!dt{{\langle F^i(t)F^i(0)\rangle_{\rm HQ}}}\equiv\frac{1}{3}\int_{-\infty}^{+\infty}\!\!dt\; D^>(t).
\eeq
In the above the expectation value is taken over a thermal bath of gluons and light quarks plus a (infinitely) heavy quark frozen to its position: the problem is reduced to evaluate a force-force correlator within this ensemble. In particular, exploiting the KMS boundary conditions obeyed by bosonic correlators, $\kappa$ is given by the zero-frequency limit of the corresponding spectral density $\sigma(\omega)$:
\beq
{\kappa}\equiv\lim_{\omega\to 0}\frac{{D^>(\omega)}}{3}\equiv\lim_{\omega\to 0}\frac{1}{3}{\frac{\sigma(\omega)}{1-e^{-\beta\omega}}}\underset{\omega\to 0}{\sim}\frac{1}{3}\frac{T}{{\omega}}{\sigma(\omega)}.
\eeq
For static heavy quarks the only force playing a role is the chromo-electric field and lattice calculations have to extract the above spectral function from the knowledge of the euclidean electric-field correlator $D_E(\tau)$ for a limited set of values of $\{\tau_i\}$: this represents a challenging issue, always encountered when attempting to get information on real-time quantities from l-QCD simulations performed in an euclidean space-time. First results have started appearing in the literature~\cite{lat1,lat2} in the case of a pure gluon plasma and work is in progress to extrapolate them to the continuum limit~\cite{lat3}: current results are shown in Fig.~\ref{fig:lQCD}

A first important message one can extract from the above discussion is that beauty measurements in heavy-ion collisions in the next years will allow one to establish a {link between} first-principle {theoretical predictions} (lattice-QCD calculations) and experimental observables: first of all the condition $M\!\gg\!gT$ makes the Langevin equation equivalent to the Boltzmann equation; secondly, being also $M\!\gg\! T$, static ($M\!=\!\infty$) l-QCD results are more reliable for beauty. This challenge will require both theoretical (in providing continuum extrapolated lQCD results including also light dynamical quarks in the simulations) and experimental (in particular in the upgrade of the detectors~\cite{upgrade}) efforts, so to make direct B measurements possible (experimental data being so far limited to the non-prompt $J/\psi$'s from B decays~\cite{CMS}).

\begin{figure}
\begin{center}
\includegraphics*[width=0.47\textwidth]{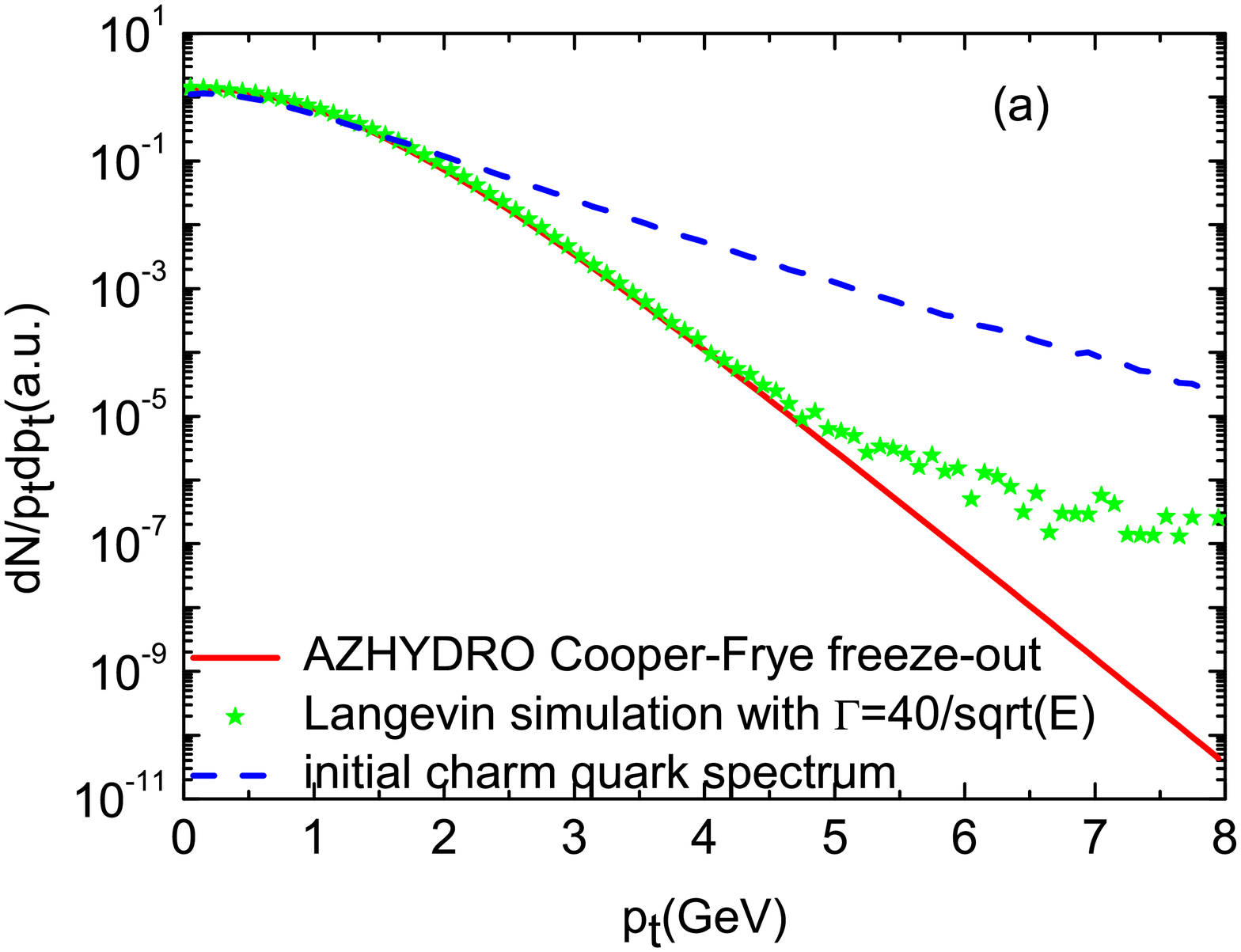}
\includegraphics*[width=0.47\textwidth]{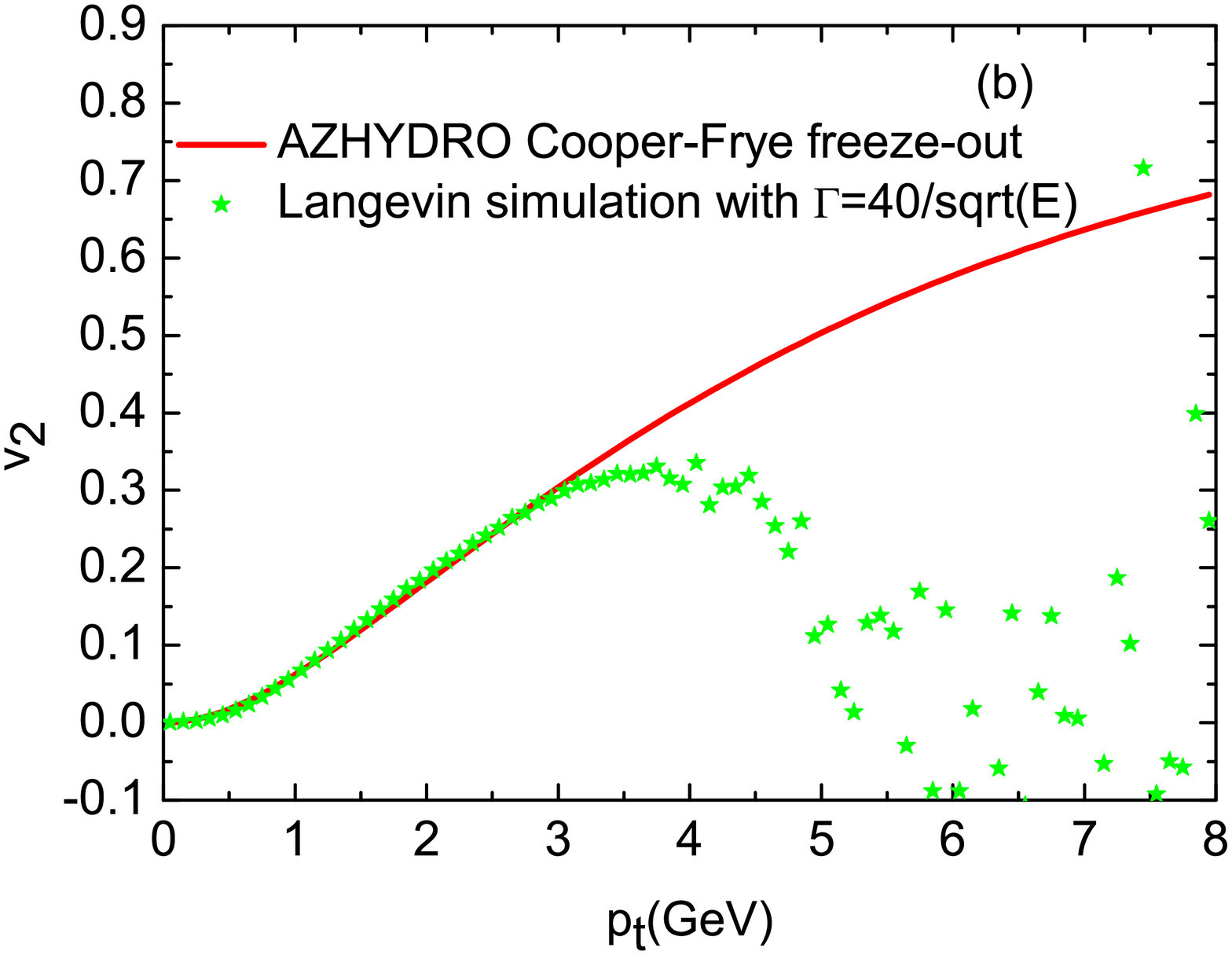}
\caption{A test of the relaxation to thermal equilibrium of charm quarks in an expanding fireball. The outcomes of Langevin simulations with very large transport coefficients display a perfect overlap with the Cooper-Frye spectra up to quite high $p_T$.}
\label{fig:validation}
\end{center}
\end{figure}

Having illustrated the tools to describe the heavy quark propagation in the fireball and shown how their transport coefficients, encoding their interaction with the medium, can be derived in principle from theory it is now possible to address the issue of heavy quark thermalization in heavy ion collisions. This entails a number of related questions. First of all one may wonder whether the above mentioned calculations would be able to describe the relaxation to thermal equilibrium in the real case of an expanding fireball, when the interaction with the medium is strong enough. If so the spectra of heavy particles decoupling at the freeze-out temperature $T_{\rm FO}$ would be given by the same Cooper-Frye algorithm employed for light hadrons, namely
\beq
E(dN/d^3p)=g\int_{\Sigma_{\rm FO}} \frac{p^\mu\!\cdot\!d\Sigma_{\mu}}{(2\pi)^3}\,\exp[-p\!\cdot\! u/T_{\rm FO}].
\eeq  
Such an issue was studied in detail in~\cite{validation}, where the propagation of $c$ quarks throughout a fireball reproducing semi-central Au-Au collisions at RHIC was simulated through a Langevin equation with very large transport coefficients. Results are shown in Fig.~\ref{fig:validation}: as it can be seen, Langevin results nicely fall on top of the Cooper-Frye curves, providing a non-trivial validation of the transport calculations. 

\begin{figure}
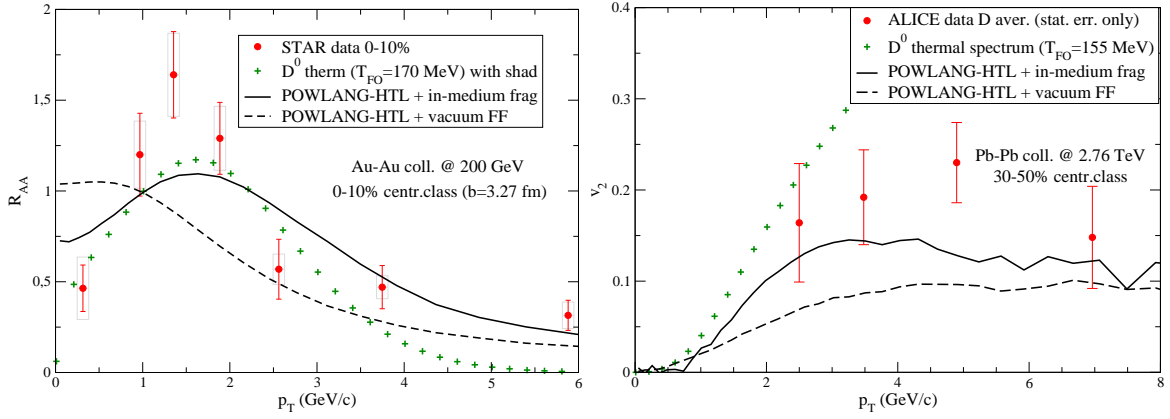

\begin{center}
\includegraphics*[width=0.46\textwidth]{RAA_D0_0-10_POW+string_nPDF.eps}
\includegraphics*[width=0.46\textwidth]{v2_D0_LHC2_POW+stringvsvac.eps}
\caption{POWLANG predictions for the $R_{AA}$ and $v_2$ of $D$ mesons at RHIC and LHC energies compared to STAR~\cite{STAR} and ALICE~\cite{v2} data. The modeling of hadronization has a strong effect, the additional flow inherited from the light thermal partons leading to a closer agreement with the data in the case of in-medium fragmentation. Also shown is the kinetic equilibrium limit.}
\label{fig:POWLANG+SF}
\end{center}
\end{figure}

A second important question is whether final \emph{hadronic} observables are able to provide unambiguous information on (possible) charm thermalization. As it can be seen in Fig.~\ref{fig:POWLANG+SF} some features of the experimental data, like the peak in the $R_{AA}$ of $D$ mesons measured by STAR and the the sizable elliptic flow observed in semi-peripheral collisions by ALICE~\cite{v2}, seem to suggest the possibility for $D$ meson to reach thermal equilibrium (Cooper-Frye spectra, green crosses) with the fireball: with charm following the flow of the medium, low-$p_T$ spectra would be boosted by radial flow and the large $v_2$ would reflect the azimuthal anisotropy of the expansion. However this is not the only possible interpretation of the data. Besides the rescattering of $c$ quarks in the plasma, $D$ meson spectra may be affected by the modification of hadronization in the presence of the medium (with its hydrodynamic expansion). It was already mentioned how transport calculations including charm coalescence with light thermal partons succeed in reproducing the STAR $R_{AA}$. Here we wish to illustrate the effect of different hadronization models interfaced to the same transport calculation in the QGP phase. For the latter we employed our POWLANG code. In Fig.~\ref{fig:POWLANG+SF} one can see how, with standard in-vacuum fragmentation of $c$ quarks, theory curves for $D$ mesons are not able to reproduce the peak in the $R_{AA}$ found by STAR and largely underpredict the $v_2$ measured by ALICE. We have then developed the following hadronization routine~\cite{inprogress}, based on the Lund string fragmentation model: at {$T_{\rm FO}$} {c quarks are coupled to light antiquarks $\overline{q}$'s} from a local {thermal distribution}, eventually {boosted} ({$u^\mu_{\rm fluid}\!\ne\!0$}) to the lab frame; {strings are then formed} and given to PYTHIA 6.4 to simulate their fragmentation and produce the final hadrons ($D+\pi+\dots$). One can see how the additional radial and elliptic flow provided by the light thermal partons move theory results (continuous curves) closer to the experimental points. The interpretation of the experimental results is then not conclusive. In particular more data at low $p_T$, so far out of reach, would be necessary at the LHC and would help to dissipate ambiguities: this will become possible with the upgrade program of the ALICE detector.

\begin{figure}
\begin{center}
\includegraphics*[width=0.46\textwidth]{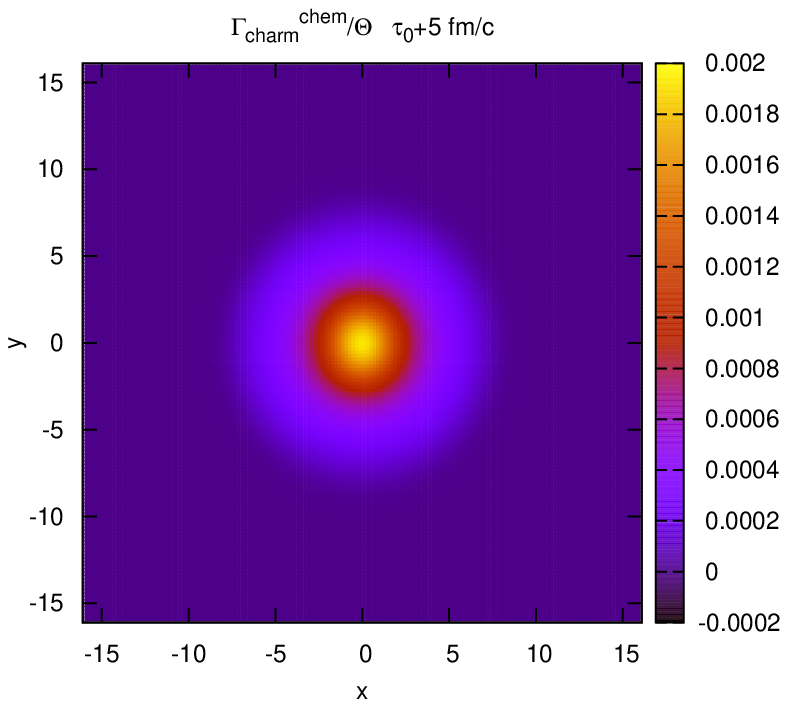}
\includegraphics*[width=0.46\textwidth]{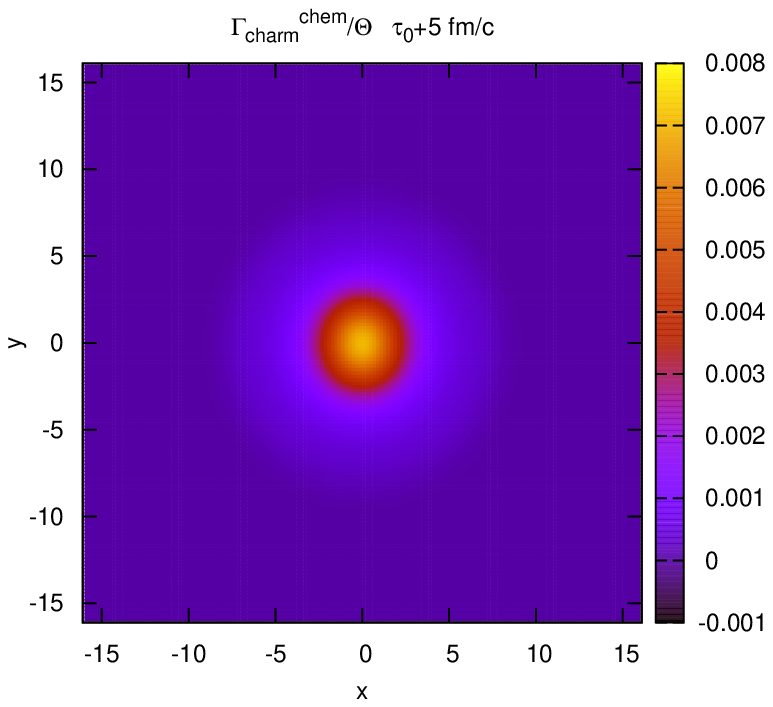}
\caption{The ratio between the charm chemical equilibration rate and the expansion rate of the fireball after 5 fm/c of evolution, obtained with typical LHC ($\sqrt{s_{\rm NN}}=2.76$ TeV, left panel) and conceivable FCC ($\sqrt{s_{\rm NN}}=39$ TeV, right panel) initial conditions in central Pb-Pb events.}
\label{fig:chemical_eq}
\end{center}
\end{figure}

While the above discussion was focused on the possibility for charm to approach kinetic equilibrium with the plasma, assuming that $c$ quarks are produced at the beginning and that their number is conserved during the limited lifetime of the medium, people have recently started wondering whether a scenario of chemical equilibrium of charm might be conceivable in the future. This is motivated both by the ongoing discussion on future experimental facilities (like the FCC, reaching a center-of-mass energy {$\sqrt{s_{NN}}\!=\!39$ TeV}~\cite{FCC}) and by recent lattice results on the contribution of charm to the QCD thermodynamics~\cite{ctherm}: the latter starts playing a non negligible role already for experimentally accessible temperatures. However the question is whether, in the actual experimental situation, with an expanding fireball with a limited lifetime $\sim 10$ fm/c, $c$ quarks have time to reach full thermal and chemical equilibrium with the rest of the medium so to be considered part of the plasma on the same footing as light quarks. Such an issue was addressed in~\cite{laine}, where the heavy quark chemical equilibration rate, starting from its perturbative result, was recast in a form suited to l-QCD simulations:
\beq
\Gamma_{\rm chem}=\frac{g^4C_F}{8\pi M^2}\left(2C_F-\frac{N_c}{2}+N_f\right)\left(\frac{TM}{2\pi}\right)^\frac{3}{2}e^{-M/T}=\frac{2\pi\alpha_s^2T^3}{9M^2}\left(\frac{7}{6}+N_f\right){\frac{\chi_{c}}{\chi_{\rm light}}},
\eeq
$\chi_i$ being the Quark Number Susceptibility of flavor $i$. Lattice results~\cite{latQNS} lead then to the estimates $\Gamma_{\rm chem}^{-1}\!\sim\! 60$ fm/c for $T\!\sim\! 400$ MeV and $\Gamma_{\rm chem}^{-1}\!\sim\! 10$ fm/c for $T\!\sim\! 600$ MeV, representing typical values of initial temperature at LHC and (conceivable) FCC Pb-Pb collisions. These numbers have to be compared not only to the expected lifetime of the produced fireball, but also to its expansion rate $\theta\!\equiv\!\partial_\mu u_{\rm fluid}^\mu$. In Fig.~\ref{fig:chemical_eq} we display two snapshots of the ratio $\Gamma_{\rm chem}/\theta$ obtained with the ECHO-QGP~\cite{ECHO} hydrodynamic code. Bearing in mind that the estimates are obtained with the analytic perturbative result for $\Gamma_{\rm chem}$, it looks that {charm} remains {far from chemical equilibrium} both at LHC ($e_0\sim 100$ GeV/fm$^3$) and at possible FCC energies ($e_0\sim 250$ GeV/fm$^3$).
 
\section{Quenching of heavy flavor jets}
\begin{figure}
\begin{center}
\includegraphics*[width=0.7\textwidth]{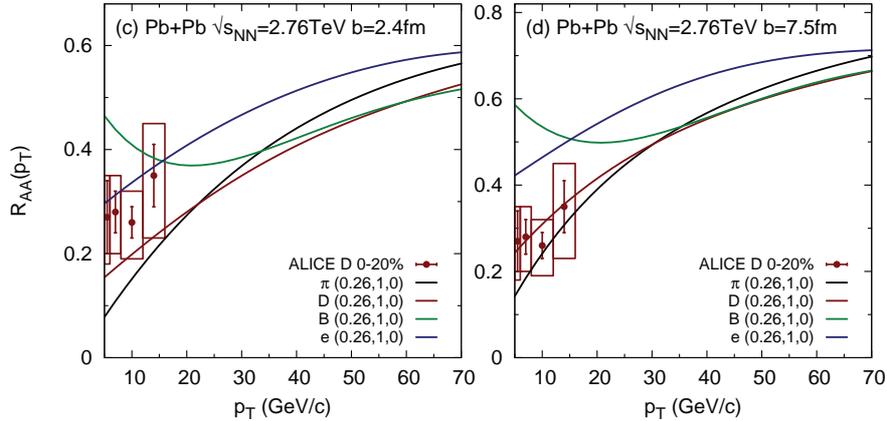}
\caption{The $R_{AA}$ of various high-$p_T$ particles in Pb-Pb collisions at the LHC predicted by the CUJET2.0 code of Ref.~\cite{gyulassy} compared to ALICE data for $D$ mesons in the highest $p_T$ bins so far accessible.}
\label{fig:highpT}
\end{center}
\end{figure}

Heavy flavor particles at high $p_T$ play a different role as probes of the medium formed in high energy nuclear collisions. They are used to tag jets, becoming then a tool to study the color charge (affecting the rate of gluon emission) and mass (modifying the formation time and the angular distribution of the radiated gluons) dependence of parton (radiative) energy loss. From the theory side one can estimate the above effects already at the lowest order of an opacity expansion in a homogeneous medium of length $L$. The spectrum of radiated gluons, here written for the case of a medium with dynamic scattering centers~\cite{magda}, reads ($x$ being the fraction of light-cone energy $E^+\!\equiv\!E\!+\!p^z$ and $\k$ the transverse momentum carried away by the radiated gluon)
\beq
{x\frac{dN_g}{dxd\k}}=\frac{C_F\alpha_s}{\pi^2}\left(\frac{L}{{\lambda_{\rm dyn}}}\right)\int d\q{\frac{\mu^2}{\pi\,{\q^2}(\q^2+\mu^2)}}\left[1-
\frac{\sin\left(\frac{(\k\!-\!\q)^2\!+\!{\chi^2}}{xE^+}L\right)}{ \frac{(\k\!-\!\q)^2\!+\!{\chi^2}}{xE^+}L }
\right]
\frac{-2(\k-\q)}{(\k-\q)^2+{\chi^2}}\left(\frac{\k}{\k^2+{\chi^2}}-\frac{\k-\q}{(\k-\q)^2+{\chi^2}}\right).
\eeq
In the above the overall Casimir factor $C_F$ reflects the color charge of the projectile and the heavy quark mass $M$ enters through the quantity $\chi^2\!\equiv\! x^2M^2\!+\!m_{\rm gluon}^2$, affecting both the gluon formation time $\tau_f\!\sim\! xE^+/[(\k\!-\!\q)^2\!+\!\chi^2]$ and the angular distribution of the radiation. While most of jet-quenching studies in the literature provide a quite rough picture of the medium, in terms of an ensemble of static scattering centers sources of Yukawa-like potentials, the above equation refers to the realistic case of a thermal bath of gluon and light quarks. The main effect is the appearance of unscreened magnetic interactions, leading to replace the simple Debye-screened potential and gluon mean free path by
\beq
{\frac{\mu^2}{\pi\,(\q^2+\mu^2)^2}}\quad\longrightarrow\quad{\frac{\mu^2}{\pi\,{\q^2}(\q^2+\mu^2)}}\quad\quad \lambda_g\quad\longrightarrow\quad\lambda_{\rm dyn}\equiv [3\alpha_sT]^{-1}.
\eeq
The dynamic modeling of the medium enhances the cross section for soft momentum exchange with respect to the static case and leads to an enhancement of the radiative energy loss. The above ideas have been recently implemented into a Monte Carlo code~\cite{gyulassy} allowing one to face the complexity of the experimental situation in which high-energy partons -- produced in initial hard processes -- cross the fireball formed in the collision, for which a realistic description based on hydrodynamics is employed: results are shown in Fig.~\ref{fig:highpT}
 
From the experimental side quenching of b-tagged jets has been recently measured by CMS~\cite{btag}, showing a suppression compatible with the one found for light flavors.

\section{Conclusions}
We have illustrated the role of heavy flavor particles as probes of the properties of the medium formed in heavy ion collisions. We have stressed the importance of future experimental data at low $p_T$ and of direct $B$ measurements to answer questions on the possibility of heavy flavor thermalization and to put tight constraints on the corresponding transport coefficients, with a solid theory-to-experiment comparison.

\section*{Acknowledgments}
I would like to thank my collaborators A. De Pace, M. Monteno, M. Nardi and F. Prino for the constant interaction.








\end{document}